\documentclass[11pt,a4paper]{article}

\usepackage[utf8]{inputenc}
\usepackage[T1]{fontenc}
\usepackage[margin=1in]{geometry}
\usepackage{graphicx}
\usepackage{hyperref}
\usepackage{amsmath,amssymb}
\usepackage{booktabs}
\usepackage[numbers,sort&compress]{natbib}
\usepackage{url}
\usepackage{xcolor}
\usepackage{authblk}
\usepackage{abstract}
\usepackage{tabularx}
\usepackage{array}

\hypersetup{
    colorlinks=true,
    linkcolor=blue,
    citecolor=blue,
    urlcolor=blue
}

\title{\textbf{Cyberlanguage: Native Communication for the Cyber-Physical-Social-Thinking Fusion Space}}

\author[1]{Huansheng Ning\thanks{Corresponding author: \href{mailto:ninghuansheng@ustb.edu.cn}{ninghuansheng@ustb.edu.cn}}}
\author[2]{Jianguo Ding}
\affil[1]{\textit{University of Science and Technology Beijing} \\ \texttt{ninghuansheng@ustb.edu.cn}}
\affil[2]{\textit{Blekinge Institute of Technology} \\ \texttt{jianguo.ding@bth.se}}

\date{}

\begin{document}

\maketitle

\begin{abstract}
Human communication is undergoing a fundamental paradigm shift. Physical space, social relations, mental states, and digital information are converging into a unified cyber-physical-social-thinking (CPST) fusion space, rendering them no longer separable domains. However, all existing communication systems, including natural and programming languages, as well as interaction protocols, were designed for a world in which these four dimensions remained distinct. We introduce Cyberlanguage, a theoretically grounded communicative framework that is native to the CPST fusion space. Grounded in the philosophical orientation of cyberism and employing CPST theory as an analytical framework, Cyberlanguage possesses four core characteristics: native four-dimensional fusion, multi-agent universality, dynamic compilability, and contextual adaptability. We have constructed a semiotic model based on the Cybersign unit, a four-dimensional synchronous grammar, a five-layer architectural stack, and a context-driven pragmatic mechanism. We also present testable empirical predictions and a staged implementation roadmap. Cyberlanguage is not intended to replace natural or programming languages, but rather to serve as a meta-communication infrastructure capable of coordinating heterogeneous agents---humans, artificial intelligences, robots, and digital entities---within an increasingly fused cyber--physical--social--cognitive reality.
\end{abstract}

\noindent\textbf{Keywords:} Cyberlanguage; Cyberism; CPST theory; heterogeneous agents; semiotics; human--computer interaction; communication paradigms

\section{The CPST Dilemma: Why Communication Needs Four Dimensions}

Communication is the foundational substrate of human civilization. Each revolution in communicative technology, from spoken language to digital networks, has reshaped social structures and modes of collective sense-making~\cite{wiener1961,gerovitch2002,ning2026cyberism,floridi2014,shannon1948}. Yet the present moment confronts us with a qualitatively novel challenge: the ontological structure of communication itself is shifting. Objects, contexts, and agents are no longer confined to a single dimension.

Consider a near-future scenario: a crisis-management coordinator issues commands that simultaneously direct drone swarms navigating physical space, deploy large-scale emergency-response robot fleets across disaster zones, activate human emergency-response teams embedded in social networks, update decision-support AI models running cognitive simulations, and query real-time data streams from a digital-twin urban model. Within a single communicative act, the word ``danger'' simultaneously indexes physical obstacles, social opinion risks, cognitive judgment biases, and algorithmic anomalies.

Existing emergency communication systems offer only partial solutions. Internationally, standardized emergency languages---brief, codified signals such as SOS, Mayday, and Pan-Pan---have long been developed to overcome cross-linguistic barriers in crisis situations. These systems are effective within their respective application domains (maritime, aviation, medical triage, etc.), but each operates within a single semiotic layer, coordinating human-to-human action in physical space. They were never designed to simultaneously address machine agents, cognitive models, and algorithmic processes across heterogeneous operational environments. No existing communication framework can integrate these multi-layered referential threads---spanning physical, social, cognitive, and computational domains---into a coherent and computable whole.

This limitation reflects a structural mismatch: the four dimensions of physical reality~(P), social organization~(S), thinking and cognition~(T), and digital cyber-infrastructure~(C) are progressively fusing into what we term the CPST fusion space~\cite{ning2025cybersophy,ning2023cyberology,ning2015cpst,shi2025}, yet our symbolic communication systems remain anchored to an era in which these dimensions remained separate. Natural languages capture P, S, and T with expressiveness but remain computationally opaque; programming languages cover C effectively but cannot natively represent social norms or physical referents.

This paper proposes Cyberlanguage as the native communication system of the CPST fusion space: a theoretically principled framework capable of (i)~simultaneously encoding referents across all four CPST dimensions; (ii)~being used equally by human, AI, robotic, and digital-human agents; and (iii)~dynamically adapting its semantic interpretation to real-time four-dimensional context. We articulate the philosophical foundations of this proposal, its architectural components, its operational mechanisms, and a set of testable empirical predictions to evaluate and refine the framework. This paper is positioned as a Perspective: a theoretically grounded conceptual framework explicitly designed to be open to critique, testable, and capable of guiding future empirical and technical research.

\section{Theoretical Foundations}

\subsection{Cyberism: A New Ontological Orientation}

The philosophical lineage of Cyberlanguage begins with Wiener's cybernetics, which sought a universal language capable of describing machines, organisms, and social systems uniformly~\cite{wiener1961,stiegler1998}. Yet classical cybernetics remained dualistic: it described organisms and machines communicating across an informational interface, rather than positing their ontological co-constitution.

The philosophical orientation we term Cyberism carries a stronger ontological commitment~\cite{ning2026cyberism,floridi2014,austin1962}: it asserts that the physical, social, thinking, and cyber dimensions are not merely related but mutually constitutive---no dimension can be fully described without reference to the others. This is neither technological determinism nor digital idealism, but a form of relational realism: any entity exists within and is partly constituted by its four-dimensional relational network.

Cyberism rests on three core propositions:

\textit{The Four-Dimensional Co-existence Proposition}: Any human experience or agent action simultaneously possesses physicality, sociality, cognitivity, and digitality. There are no pure ``physical facts'' or ``digital facts'' independent of the other three dimensions.

\textit{The Relation-Primacy Proposition}: An entity's identity is constituted by its four-dimensional relational network, not by intrinsic attributes. An ``object'' is what it is because of its physical composition, social meaning, cognitive representation, and digital instantiation simultaneously.

\textit{The Communication-Constitutive Proposition}: Communicative acts in the CPST space do not merely transmit pre-formed information; they generate and transform four-dimensional relations through fundamental carriers and protocols that operate across all four dimensions. Communication is an ontological practice, not merely an informational one~\cite{austin1962,searle1969}.

\subsection{CPST Theory: A Four-Dimensional Analytical Framework}

The CPST framework provides methodological architecture for operationalizing Cyberism in communication analysis. The four dimensions are:

\textit{Physical}~(P): The natural world and its regularities---objects, spatial relations, temporal sequences, and causal structures~\cite{gibson1979,gibson2015}.

\textit{Social}~(S): Networks of human relations and their institutional forms---roles, norms, rights, obligations, and trust, as well as emergent collective intelligence and social dynamics~\cite{latour2005,castells1996,thrift2005}.

\textit{Thinking}~(T): Mental states and cognitive processes---intentions, beliefs, desires, attention states, and epistemic attitudes~\cite{hutchins1995,bateson2000}.

\textit{Cyber}~(C): Digital information space and its computational processes---data structures, algorithms, machine-learning models, network topologies, and computational states~\cite{kitchin2011,hui2016}.

The critical insight of CPST theory is that these four dimensions are mutually constituting rather than independently stacked. Any communicative event can be formally located within this CPST quadspace---a term we adopt to denote the jointly constituted four-aspect analytical space, distinct from the geometric or physical notion of dimensionality---in which meaning is the joint product of all four coordinates simultaneously. As shown in Table~\ref{tab:cpst_dimensions}, existing communication systems cover subsets of this space, but none cover all four natively.

\begin{table}[htbp]
\centering
\caption{The CPST Four-Dimensional Framework: Dimensions, Core Elements, Communicative Functions, and Limitations of Existing Systems}
\label{tab:cpst_dimensions}
\small
\renewcommand{\arraystretch}{1.3}
\begin{tabular}{@{}p{0.12\textwidth} p{0.22\textwidth} p{0.28\textwidth} p{0.32\textwidth}@{}}
\toprule
\textbf{Dimension} & \textbf{Core Elements} & \textbf{Communicative Functions} & \textbf{Limitations in Existing Systems} \\
\midrule
Physical (P) & Objects, space, time, causality, material states & ``The device is at location X''; physical event triggers & Natural language handles well but ambiguously; no machine-readable physical ontology \\
Social (S) & Roles, norms, rights, trust, institutional context & ``You are authorized''; social obligation declarations & Requires social commonsense absent in machines; role ambiguity across cultures \\
Thinking (T) & Intentions, beliefs, emotions, attention, epistemic states & ``I believe X''; goal-directed instructions with confidence levels & Mental predicates computationally opaque; subjective states non-standardized \\
Cyber (C) & Data, algorithms, models, computation, network states & ``Query database Y''; algorithm invocation; digital twin state updates & Programming languages cover C well but remain isolated from P, S, T \\
\bottomrule
\end{tabular}
\renewcommand{\arraystretch}{1.0}
\end{table}

\subsection{Historical Evolution of Communication Paradigms}

The history of communication media can be reframed as the progressive integration of CPST dimensions. Five paradigmatic phases are identifiable, each characterized by a distinctive dimensional configuration:

\textbf{Phase~1 --- Oral Era (P+S+T):} Face-to-face communication integrates physical co-presence, social relations, and mental states, but lacks an independent cyber dimension. Information storage depends on embodied memory.

\textbf{Phase~2 --- Written Era (P+S+T + nascent C):} Writing allows information to exist independently of the speaker, initiating a proto-digitization process. Carriers remain physical; computational capacity is absent.

\textbf{Phase~3 --- Mass Media Era (P+S+T + one-way C):} Print, broadcast, and television enable large-scale reproduction of information. Interaction remains one-directional; the cyber dimension is operational but non-reciprocal~\cite{mcluhan1994,kittler1999,negroponte1995,parikka2012}.

\textbf{Phase~4 --- Digital Media Era (P+S+T + bidirectional C):} The Internet grants the cyber dimension parity with physical space for the first time~\cite{gibson1979,castells1996,bratton2015,benkler2006,mitchell1995}. The four dimensions coexist but remain ontologically ``parallel'': the virtual/real and online/offline distinctions remain operative.

\textbf{Phase~5 --- CPST Fusion Era:} The Internet of Things datafies physical environments; social networks are algorithmically mediated; brain--computer interfaces render mental states computationally tractable; digital twins model physical and social systems in real time~\cite{ning2025cybersophy,ning2023cyberology,willett2023,metzger2023,rasheed2020}. The four dimensions are no longer parallel layers but mutually interpenetrating~\cite{stiegler1998,clark2003,hayles1999,haraway1991,hansen2006}. Cyberlanguage is the communicative response to this ontological condition.

\begin{figure}[htbp]
    \centering
    \includegraphics[width=1.05\textwidth]{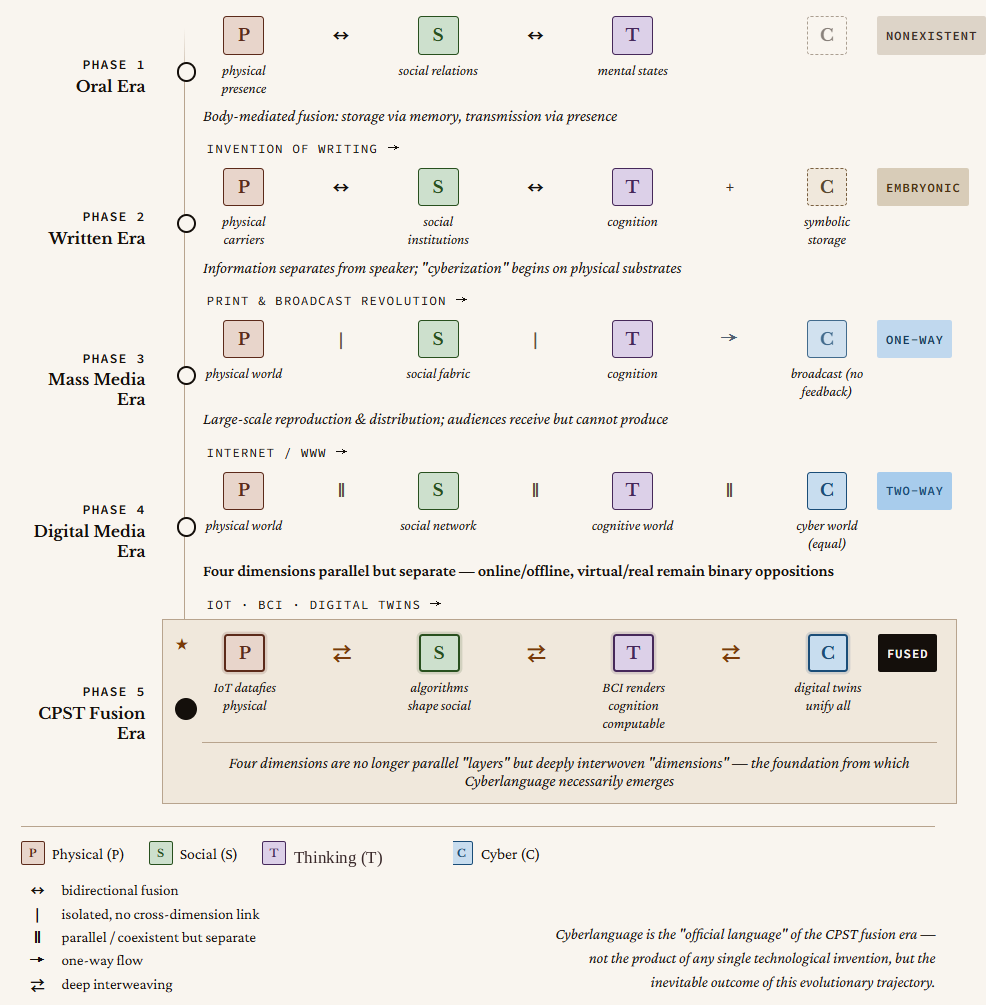}
    \caption{Historical evolution of the four communication dimensions (Physical, Social, Thinking, Cyber) across five eras. Each phase is characterised by the structural relationship between dimensions and the developmental status of the Cyber~(C) domain, ranging from its complete absence in the Oral Era to its constitutive role in the emerging CPST Fusion Era.}
    \label{fig:evolution}
\end{figure}

\section{Cyberlanguage: Definition and Core Characteristics}

\subsection{Formal Definition}

Drawing on the theoretical foundations established above, we offer the following definition:

\begin{quote}
\textit{Cyberlanguage is a symbolic system designed for multi-dimensional communication among heterogeneous agents---including humans, artificial intelligence, robots, digital humans, and other cyber-enabled physical entities---within the CPST fusion space. It is natively compatible with the physical, social, thinking, and cyber dimensions and can achieve consistent semantic mapping and dynamic reconstruction across all four dimensions in real time.}
\end{quote}

This definition has three key components. First, Cyberlanguage is defined over the entire CPST quadspace rather than being confined to any single constituent aspect. Second, its user scope extends beyond humans to encompass heterogeneous agents---departing from all existing communication frameworks, which presuppose at least one human participant~\cite{peirce1931,saussure1916}. Third, its functional goal is to achieve semantically consistent mapping and integration across all four CPST aspects simultaneously, rather than sequential information transfer between disjoint channels~\cite{shannon1948,clark1996}.

\subsection{Four Core Characteristics}

\textbf{Native Fusion.} Cyberlanguage is not an overlay of natural language and programming syntax, but is architecturally designed from first principles for four-dimensional compatibility. A single Cyberlanguage expression simultaneously encodes the physical characteristics of a referent~(P), its social-contextual status~(S), the communicating agent's cognitive intention toward it~(T), and the state of its digital representation in computational systems~(C). This fusion is not post-hoc or additive---it is constitutive of the basic semiotic unit.

\textbf{Multi-Agent Universality.} Existing communication systems exhibit what we term `agent-type isolation': natural language is human-exclusive; programming languages are machine-exclusive; interaction protocols are system-exclusive. Translation between these modalities requires substantial computational overhead and introduces systematic semantic loss. Cyberlanguage is designed as a universal symbolic medium enabling humans, AIs, robots, and digital entities to communicate within a shared semiotic framework, eliminating the need for pairwise translation layers~\cite{brooks1991,park2023,yao2023,zhao2023,xi2023,touvron2023,pan2024,li2023camel,wang2024}.

\textbf{Dynamic Compilability.} Cyberlanguage supports `deep semantics' that can be dynamically compiled into multiple `surface expressions' adapted to the recipient's capabilities and the current situational context. The same underlying Cyberexpression may be rendered as a natural-language sentence for a human operator, as structured JSON for a machine-learning system, as a command sequence for a robotic actuator, or as a state-update packet for a digital twin. Furthermore, with AI-mediated translation, Cyberlanguage's deep semantic layer can serve as an interlingua for real-time mutual translation across existing human natural languages---enabling, for example, a Mandarin-speaking coordinator and an English-speaking field responder to communicate through a shared Cyberexpression that compiles into each party's native language while preserving full four-aspect semantic fidelity. This compilability is principled, not arbitrary: it is governed by formal mapping functions that connect the deep semantic representation to surface-level syntactic forms~\cite{rasheed2020,xie2021,qin2022,gunduz2023}.

\textbf{Contextual Adaptability.} Cyberlanguage continuously samples real-time CPST context and adjusts semantic interpretation accordingly. A single term such as `danger' is interpreted by dynamically weighting contributions from physical sensor readings, social authority structures, inferred cognitive states of participants, and algorithmic risk assessments. This dynamic weighting is not a heuristic but a principled pragmatic mechanism described in Section~\ref{sec:pragmatics}.

\begin{figure}[htbp]
    \centering
    \includegraphics[width=0.7\textwidth]{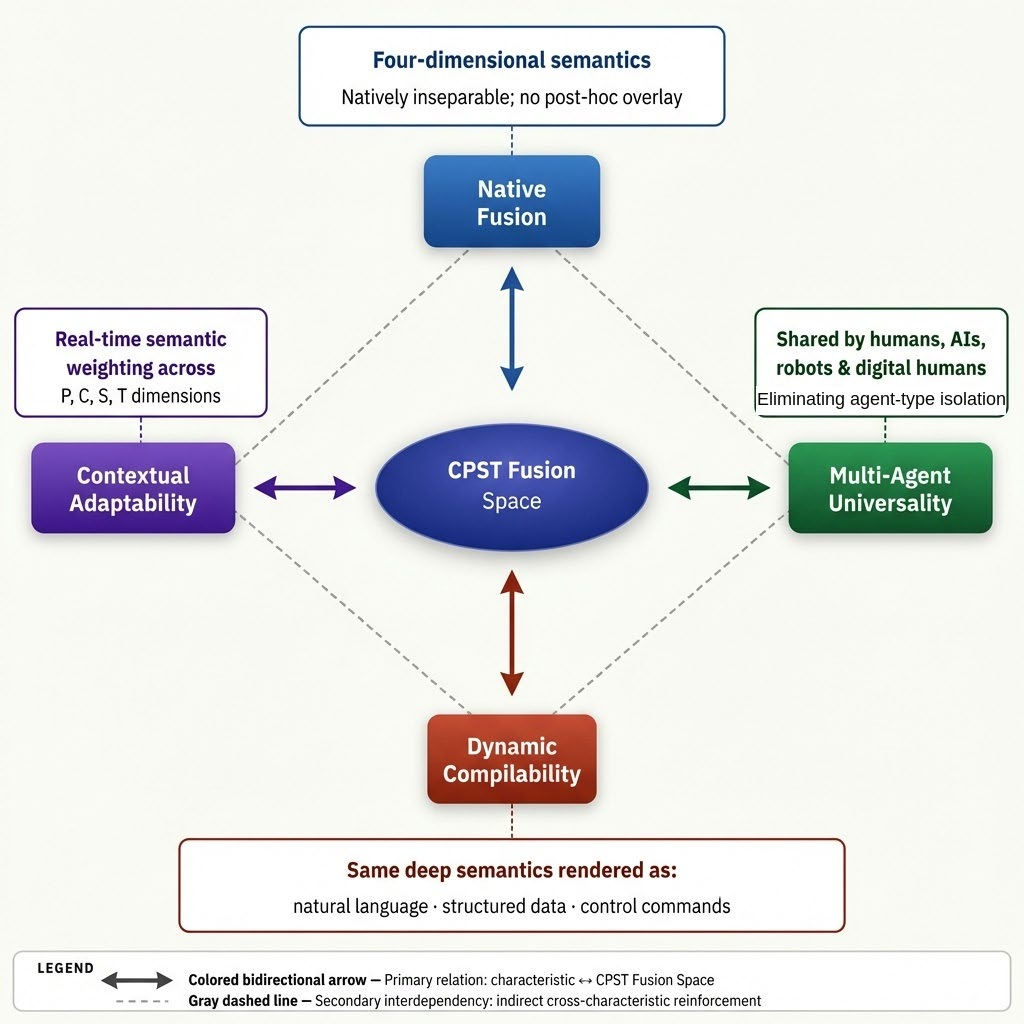}
    \caption{The four core characteristics of Cyberlanguage and their mutual interdependencies. Bidirectional arrows indicate that each characteristic presupposes and reinforces the others; the CPST Fusion Space constitutes the operational domain from which all four emerge.}
    \label{fig:characteristics}
\end{figure}

\subsection{Comparison with Existing Communication Forms}

\begin{table}[htbp]
\centering
\caption{Cyberlanguage compared with existing communication forms across six functional dimensions}
\label{tab:comparison}
\footnotesize
\renewcommand{\arraystretch}{1.3}
\begin{tabular}{@{}p{0.13\textwidth} p{0.16\textwidth} p{0.16\textwidth} p{0.16\textwidth} p{0.27\textwidth}@{}}
\toprule
\textbf{Feature} & \textbf{Natural Language} & \textbf{Programming Language} & \textbf{Interaction Protocol} & \textbf{Cyberlanguage (proposed)} \\
\midrule
Primary Users & Humans & Developers / machines & Machines / systems & Heterogeneous agents (humans, AIs, robots, digital entities, IoT devices) \\
Dimensional Coverage & P+S+T (weak~C) & C only (weak P/S/T) & C (primary) & P+S+T+C --- native four-dimensional \\
Ambiguity Tolerance & High & Extremely low & Low & Dynamically adjustable per context \\
Computability & Low & High & High & Natively computable across all four dimensions \\
Agent Interoperability & Human-to-human only & Machine-to-machine only & System-to-system only & Cross-species: human $\leftrightarrow$ AI $\leftrightarrow$ robot $\leftrightarrow$ digital twin \\
Evolution Mechanism & Organic / uncontrolled & Versioned / backward-compat. & Standards-body driven & Community consensus + governed, versioned evolution \\
\bottomrule
\end{tabular}
\renewcommand{\arraystretch}{1.0}
\end{table}

Table~\ref{tab:comparison} makes clear that Cyberlanguage does not compete with existing systems within their native domains; instead, it provides a meta-communicative substrate capable of coordinating and translating among them. Natural language, programming languages, and interaction protocols remain optimal within their respective niches; Cyberlanguage enables semantic coherence across niches when heterogeneous agents must collaborate.

\section{Architectural Framework}

\subsection{Semiotic Foundations: From Sign to Cybersign}

Saussure's~\cite{saussure1916} dyadic model (signifier/signified) and Peirce's~\cite{peirce1931} triadic model (sign, object, interpretant) each presuppose a single referential domain and cannot accommodate the simultaneous four-dimensional referential relations that characterise communication in the CPST fusion space~\cite{harnad1990,lakoff1980,wittgenstein1953}.

We introduce the Cybersign as the fundamental semiotic unit of Cyberlanguage. A Cybersign comprises a Linguistic Signifier~($\Lambda$)---the perceptible surface expression---and four-dimensional sign-relations $\Sigma_d = \langle \sigma_d, \rho_d \rangle$, each a Saussurean dyad of dimensional signifier and signified. The full structure is:
\begin{equation}
\text{Cybersign} = \langle \Lambda;\; \Sigma_C,\; \Sigma_P,\; \Sigma_S,\; \Sigma_T \rangle
\end{equation}

The four $\Sigma_d$ are interlinked through a cyber-mediated mapping architecture that reflects the infrastructural role of the Cyber dimension within Cyberlanguage. Three primary mapping functions connect each non-cyber aspect to its computational representation:
\begin{align}
f_{CP} &: \text{physical characteristics} \leftrightarrow \text{digital-twin / computational representation} \\
f_{CS} &: \text{social-contextual meaning} \leftrightarrow \text{computational encoding} \\
f_{CT} &: \text{cognitive / intentional state} \leftrightarrow \text{computational state}
\end{align}

Three derived cross-aspect mappings are then composed through the Cyber layer:
\begin{align}
f_{PS} &= f_{CS}^{-1} \circ f_{CP} : P \to C \to S \quad \text{(physical $\leftrightarrow$ social, mediated computationally)} \\
f_{PT} &= f_{CT}^{-1} \circ f_{CP} : P \to C \to T \quad \text{(physical $\leftrightarrow$ cognitive, mediated computationally)} \\
f_{ST} &= f_{CT}^{-1} \circ f_{CS} : S \to C \to T \quad \text{(social $\leftrightarrow$ cognitive, mediated computationally)}
\end{align}

Finally, the full fusion function $f_{CPST}$ synchronizes all four aspect-layers into a unified Cybersign state, ensuring four-aspect semantic coherence in any single communicative act.

This architecture does not imply that the Cyber dimension is ontologically superior to the other three; rather, it recognizes that within a computationally instantiated communication system, the Cyber layer necessarily serves as the operational substrate through which cross-aspect mappings are realized. The distinction is between implementational centrality and ontological primacy: just as a nervous system mediates between perception, cognition, and action without being ontologically prior to them, the Cyber layer mediates cross-aspect mappings without claiming ontological privilege. All six pairwise relationships remain analytically valid; the cyber-mediated structure specifies how they are implemented.

Each function operates bidirectionally---forward during encoding, in reverse during decoding---with $f_{CT}$ (cognitive/intentional state $\leftrightarrow$ computational state) the most theoretically consequential, as it formalises the mind--machine interface underlying brain--computer interaction~\cite{turing1950,clark1997,dennett1991}. Integrated Cybersign meaning~$M$ is produced by a context-sensitive inference process over all four~$\Sigma_d$ and the six~$f_{XY}$ simultaneously, as formalized in Section~\ref{sec:pragmatics}.

\begin{figure}[htbp]
    \centering
    \includegraphics[width=0.85\textwidth]{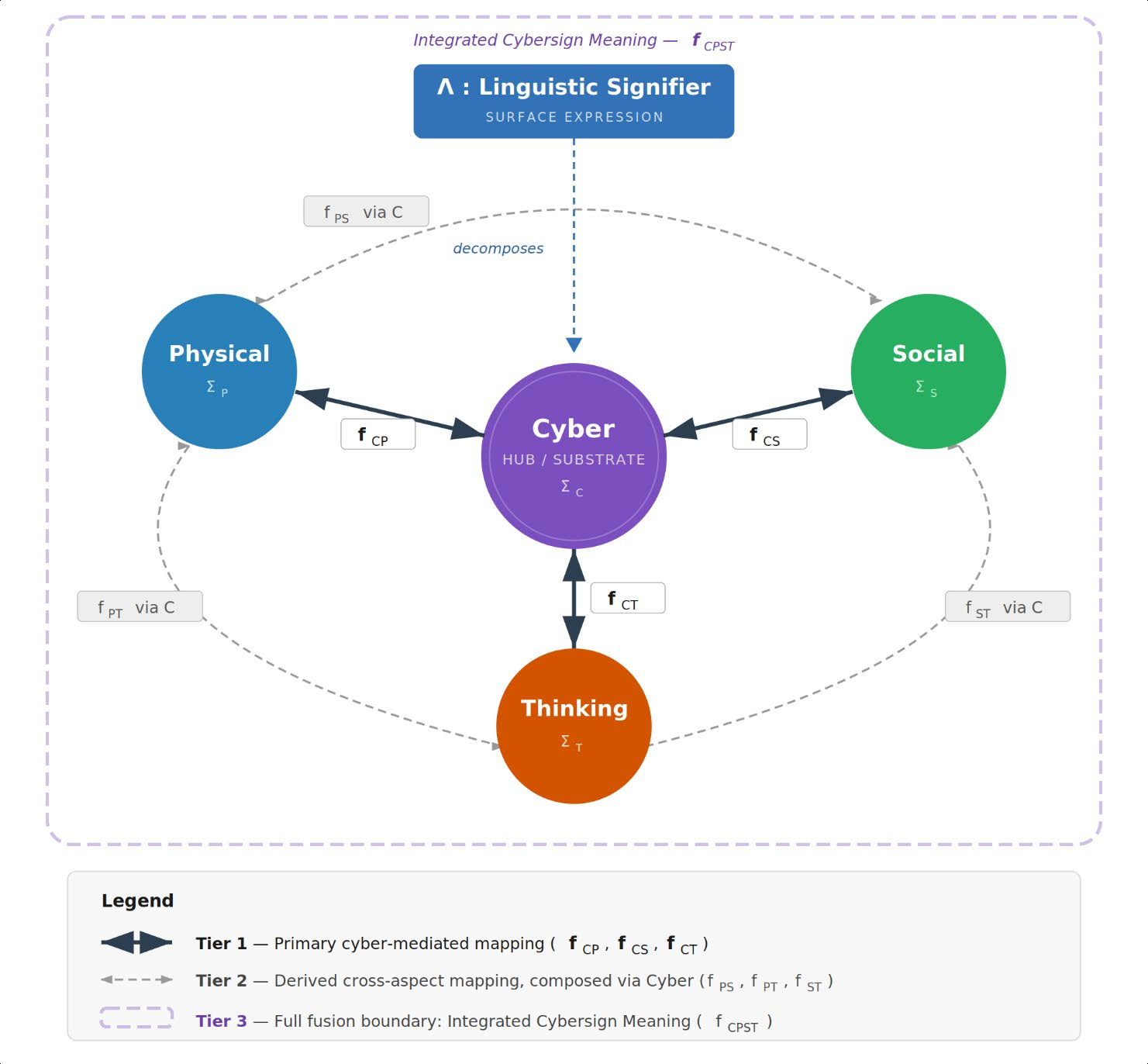}
    \caption{Four-dimensional semiotic structure of a Cybersign. The top node~($\Lambda$) is the Linguistic Signifier (surface expression form); the four coloured nodes represent the dimensional referents $\Sigma_C$ (cyber), $\Sigma_P$ (physical), $\Sigma_S$ (social), and $\Sigma_T$ (thinking). Solid arrows (Tier~1) are primary cyber-mediated mappings; dashed arcs (Tier~2) are derived cross-aspect mappings composed via the Cyber hub.}
    \label{fig:cybersign}
\end{figure}

\subsection{Grammatical Structure: Four-Dimensional Synchronous Grammar (FDSG)}

Four-Dimensional Synchronous Grammar (FDSG) is the formal syntactic system of Cyberlanguage. Its canonical Cyberstatement structure is:
\begin{equation}
[\text{P-component}] \; [\text{S-component}] \; [\text{T-component}] \; [\text{C-component}] \; \oplus \Omega
\end{equation}
where each component carries typed dimension-specific semantic content, and $\oplus\Omega$ is the Integration Operator specifying cross-dimensional composition rules (priority, parallel execution, or probabilistic blending). The FDSG formalism is grounded in dependent type theory~\cite{chomsky2002,manning1999,newell1976}.

As an illustrative example, consider a Cyberstatement for human--drone-swarm coordination in an emergency response scenario:

\smallskip
\noindent\texttt{[P: sector=A7, altitude=50m, duration=1800s]}\\
\texttt{[S: authorisation=$\alpha$, mission-id=SAR-2026-047]}\\
\texttt{[T: intent=reconnaissance, confidence=0.92, urgency=high]}\\
\texttt{[C: algorithm=path-optimize-v3, datasource=live-weather-api]}\\
\texttt{[$\oplus\Omega$: P$\succ$S, T$\|$C]}
\smallskip

\begin{figure}[htbp]
    \centering
    \includegraphics[width=0.75\textwidth]{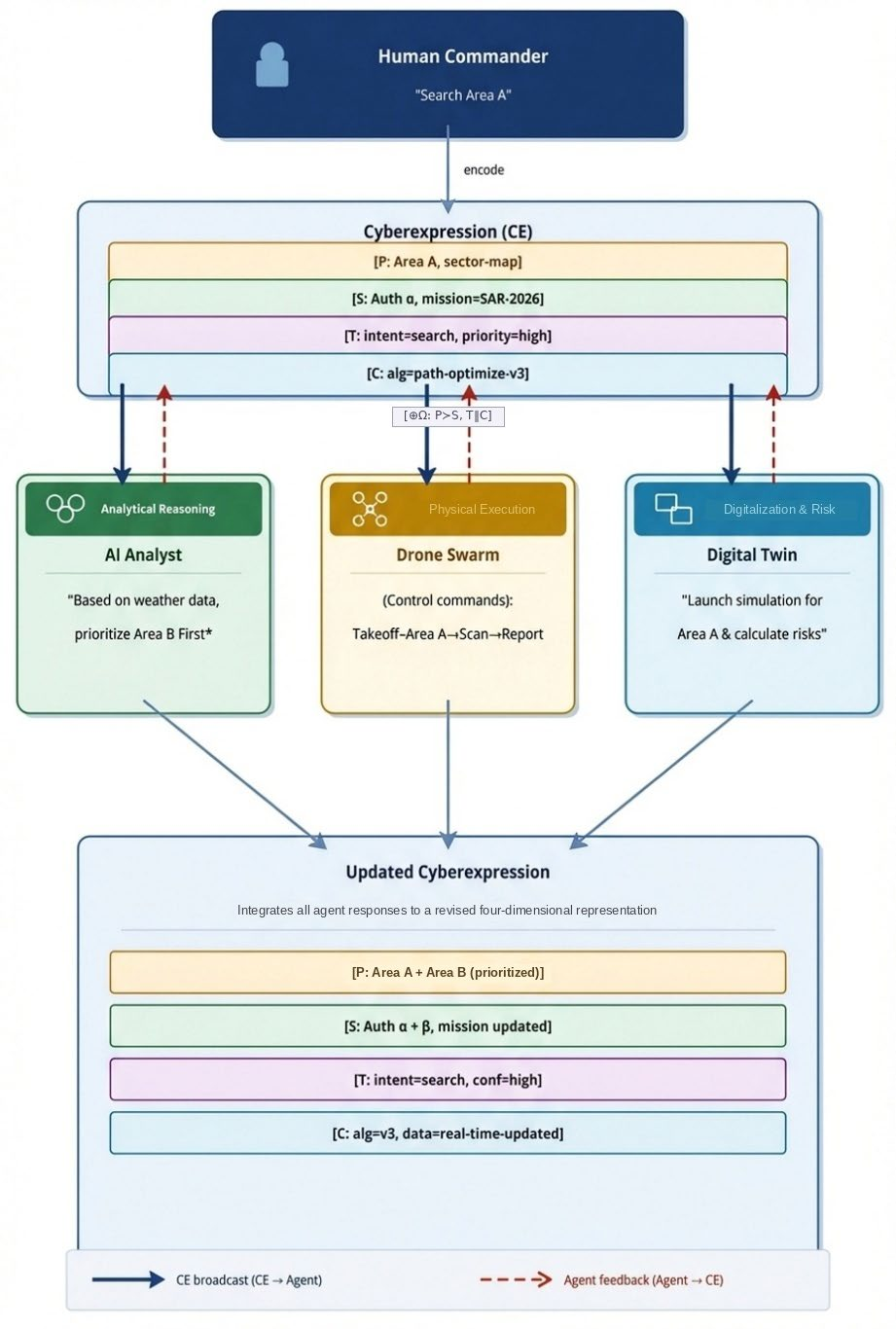}
    \caption{Cyberlanguage communication among heterogeneous agents. The Human Commander issues a natural-language instruction, encoded as a Cyberexpression (CE), and broadcasts it to three agent types: AI Analyst, Drone Swarm, and Digital Twin.}
    \label{fig:communication}
\end{figure}

The example Cyberstatement is simultaneously realisable as natural language for human commanders, a structured control command for drone swarms, a model-update event for mission AI systems, and a simulation trigger for digital-twin platforms---each recipient extracting the dimensional projection relevant to its type. Fig.~\ref{fig:communication} illustrates this process across three agent types.

\subsection{Pragmatic Mechanism: Context-Driven Meaning Generation}
\label{sec:pragmatics}

The pragmatics of Cyberlanguage extends classical speech act theory~\cite{austin1962,searle1969} and Gricean cooperative principles~\cite{grice1978} to four-dimensional contexts. For a Cyberexpression~$E$ within four-dimensional context~$C$, the Integrated Cybersign Meaning~$M$ is formalised as:
\begin{equation}
M(E, C) = \mathrm{Sem}(E) \otimes \mathrm{State}_C(C) \otimes \mathrm{State}_P(C) \otimes \mathrm{State}_S(C) \otimes \mathrm{State}_T(C)
\end{equation}
where $\mathrm{Sem}(E)$ is the inherent semantic content of the Cyberexpression, and $\mathrm{State}_C$, $\mathrm{State}_P$, $\mathrm{State}_S$, $\mathrm{State}_T$ are the real-time states of the four CPST dimensions of context~$C$. The context-integration function~$\otimes$ specifies how $\mathrm{Sem}(E)$ is weighted relative to real-time dimensional states; its precise design is a central open problem in Cyberlanguage research (Section~\ref{sec:predictions}). Candidate formalisations include learned weighted tensor products, probabilistic graphical models over the four-dimensional state variables, and attention-based neural mechanisms analogous to cross-modal fusion in multimodal transformers, each offering distinct trade-offs between interpretability and expressiveness.

\subsection{Layered Architecture}

\begin{figure}[htbp]
    \centering
    \includegraphics[width=0.75\textwidth]{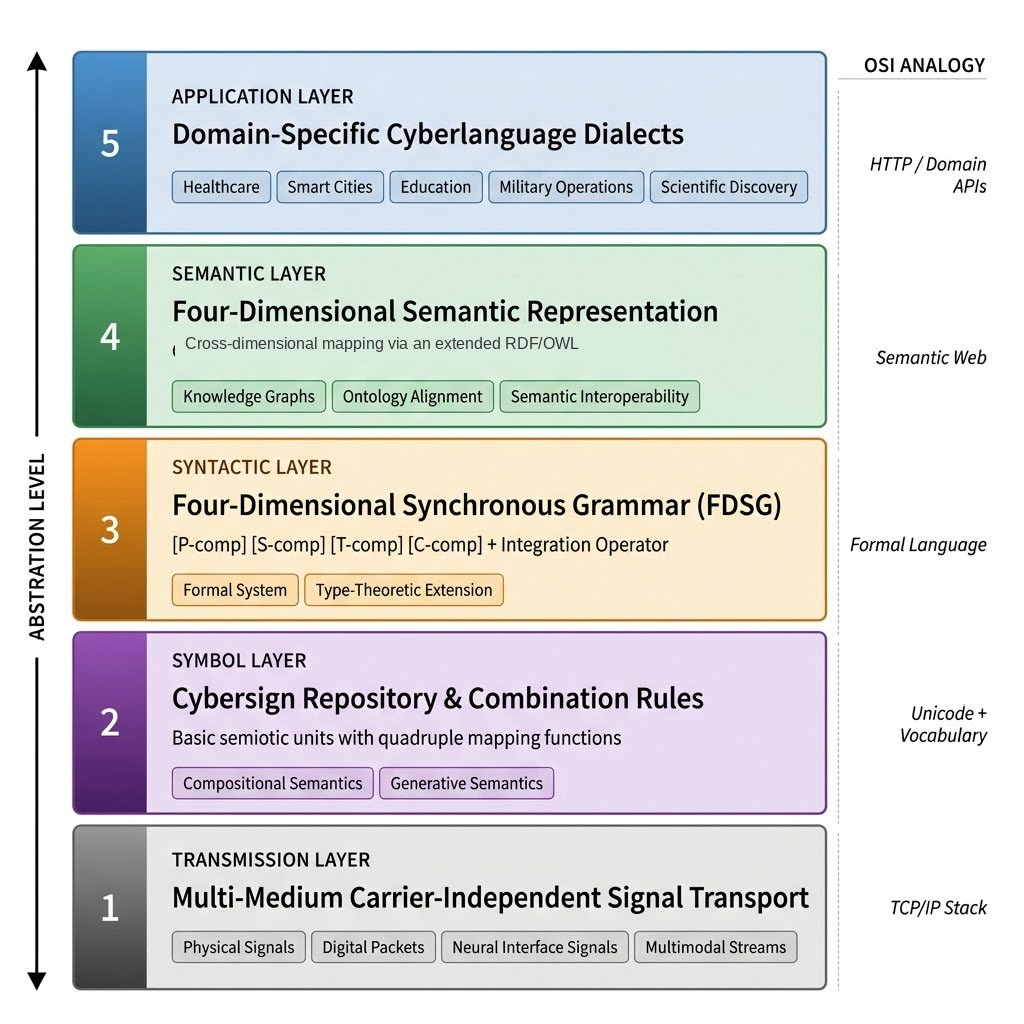}
    \caption{Five-layer architectural stack of Cyberlanguage, structured analogously to the OSI network model. Lower layers (1--2) provide carrier-independent symbolic infrastructure; upper layers (4--5) implement domain-specific communicative conventions. Layer~3 mediates via a formal grammar.}
    \label{fig:architecture}
\end{figure}

Cyberlanguage is not a monolithic system but a layered symbolic architecture, analogous to the OSI model in network communication. Fig.~\ref{fig:architecture} presents the five-layer stack.

The five layers, from bottom to top, are: Layer~1 (Transmission), providing multi-medium carrier-independent signal transport across physical, digital, and neural channels; Layer~2 (Symbol), housing the Cybersign repository and compositional combination rules; Layer~3 (Syntactic), implementing the FDSG formalism; Layer~4 (Semantic), enabling four-dimensional semantic representation via extended ontologies such as RDF/OWL, knowledge graphs, and semantic interoperability mechanisms~\cite{pan2024kg,hogan2021,bernerslee2001}; and Layer~5 (Application), supporting domain-specific Cyberlanguage dialects for healthcare, smart cities, education, and other verticals. This architecture permits Cyberlanguage to maintain foundational unity at the Cybersign repository and FDSG formalism levels, while supporting upper-layer diversity.

\section{Operational Mechanisms}

\subsection{The Encoding--Transmission--Decoding Cycle}

The complete communication cycle of Cyberlanguage comprises three interdependent phases. During Encoding, the sender decomposes a communicative intention into four-dimensional semantic components and generates a well-formed Cyberexpression in accordance with FDSG rules. During Transmission, the Cyberexpression traverses carrier channels appropriate to the agents involved---screen text, acoustic speech, network packets, or neural interface signals~\cite{shannon1948,kress2001}. During Decoding, the receiver parses the syntactic structure, integrates it with real-time context per Section~\ref{sec:pragmatics}, and produces output in its own representational format.

Heterogeneous receivers---human, AI, robotic, or digital-twin systems---extract different projections of the same Cyberexpression and feed back updates in compatible formats, without requiring bilateral translation layers. Cyberlanguage thus functions as a universal semantic bus.

\subsection{Error Handling and Meaning Negotiation}

Cyberlanguage incorporates multi-layer mechanisms for detecting and resolving communicative failures, formalised in the four-step Meaning Negotiation Protocol. Because expressions explicitly distinguish P, S, T, and C components, ambiguities can be precisely localized to specific dimensional sources rather than remaining diffuse. The system supports:

\textit{Dimensional Explicitation}: Requesting that a communicating party make explicit the dimensional component that is the source of ambiguity.

\textit{Meta-communicative Markers}: Operators allowing agents to communicate about the communication itself (``Interpret the T-component of my last expression as intent=clarification, confidence=0.6'').

\textit{Dynamic Meaning Negotiation}: Multi-round negotiation protocols through which agents iteratively converge on a shared interpretation of a Cyberexpression, drawing on the full four-dimensional context at each round.

\textit{Adaptive Learning}: Over extended interactions, Cyberlanguage-enabled systems record systematic misunderstanding patterns and adjust semantic mappings---analogous to the gradual alignment of interpretive conventions in human linguistic communities~\cite{tomasello2008}.

\section{Implementation Roadmap}

\subsection{Staged Implementation Strategy}

\textbf{Phase~1 (Years 1--3): Domain-specific Cyberlanguages.} Implement prototype CPST interaction frameworks in bounded vertical domains---such as smart-home automation, medical telemonitoring, and collaborative manufacturing---where the four-dimensional structure of communicative events is tractable and verifiable. Validate core concepts and measure efficiency improvements relative to existing protocols.

\textbf{Phase~2 (Years 3--5): Cross-domain Semantic Mapping.} Establish mappings between domain-specific Cyberlanguages; develop general-purpose Cybersign repositories and shared grammatical rules; test in larger-scale heterogeneous systems involving multiple agent types.

\textbf{Phase~3 (Years 5--10): General Cyberlanguage.} Synthesize cross-domain experience into a general Cyberlanguage specification; promote standardization through international bodies; support an open, community-governed Cyberlanguage ecosystem.

\subsection{The CyberCorpus Initiative}

Empirical development requires CyberCorpus: a multimodal interaction corpus annotated with four-dimensional labels (P, S, T, C components and their cross-dimensional mappings). Candidate data sources include human--robot task logs, smart-home interaction records, multi-agent coordination traces, and brain--computer interface experiments. Each sample requires quadruple annotation, providing the training signal for four-dimensional encoders~\cite{zhao2023,touvron2023,wittgenstein1953}.

\subsection{Context-Aware Semantic Computing Models}

The core computational substrate of Cyberlanguage requires four capabilities: four-dimensional encoders simultaneously processing physical sensor data, social-relational graphs, cognitive-state representations, and digital information streams; dynamic contextual attention mechanisms allocating resources across dimensions; cross-agent semantic alignment ensuring consistent interpretation across heterogeneous receivers; and explainability mechanisms enabling human operators to audit the dimensional inference process~\cite{yao2023,zhao2023,touvron2023,pan2024,shumailov2024,wei2022emergent,bommasani2022,achiam2023,gu2024,wei2022cot,vaswani2017,brown2020,lecun2015,silver2016,russell2020}.

\subsection{Governance and Standardization}

As a socio-technical infrastructure, Cyberlanguage requires governance addressing: technical standardization (core grammar, Cybersign repositories, transmission protocols through ISO/IEC or analogous bodies), community participation (open governance involving academia, industry, and civil society), ethical design (bias auditing of Cybersign mappings), and legal adaptation (especially for smart-contract applications where assertions may have legal force)~\cite{shi2025,levy1997,lessig1999}.

\section{Testable Predictions and Empirical Research Agenda}
\label{sec:predictions}

A theoretical framework of this scope must be empirically testable. We articulate four concrete, falsifiable predictions that can be evaluated with near-term experimental methods, and the outcome signatures that would confirm, disconfirm, or require revision of the Cyberlanguage framework (Table~\ref{tab:predictions}). The quantitative thresholds specified are provisional benchmarks informed by analogous improvements reported in multimodal fusion and structured communication research and are subject to calibration through pilot studies.

\begin{table}[htbp]
\centering
\caption{Testable predictions, proposed methods, and expected outcomes for Cyberlanguage research}
\label{tab:predictions}
\small
\renewcommand{\arraystretch}{1.3}
\begin{tabular}{@{}p{0.28\textwidth} p{0.35\textwidth} p{0.30\textwidth}@{}}
\toprule
\textbf{Testable Prediction} & \textbf{Proposed Method} & \textbf{Expected Outcome} \\
\midrule
P1: Dimensional disambiguation reduces error in human--AI task completion & Controlled experiment: participants complete collaborative tasks using standard natural language vs.\ CPST-structured prompts & $\geq$20\% reduction in miscommunication-induced errors in the CPST condition \\
P2: Four-dimensional context encoding improves intent recognition in heterogeneous-agent systems & Benchmark: extend SQuAD/MultiWOZ with CPST annotations; compare Cybersign-aware LLM vs.\ baseline & Higher F1 on intent disambiguation for CPST-annotated model (target: $>$5\% absolute) \\
P3: Cyberlanguage acquisition follows a staged developmental trajectory in humans & Longitudinal study of non-expert users learning a domain-specific Cyberlanguage prototype (smart-home context) & P-component mastered first; C-component requires most scaffolding; T reflects individual differences \\
P4: Dimensional mapping functions can be learned from multimodal corpora & Train a cross-modal neural encoder on CyberCorpus; evaluate on cross-dimensional inference tasks & Learned $f_{PT}$ and $f_{ST}$ mappings outperform unimodal baselines on held-out CPST benchmark \\
\bottomrule
\end{tabular}
\renewcommand{\arraystretch}{1.0}
\end{table}

Beyond Table~\ref{tab:predictions}, the empirical agenda encompasses cognitive studies of four-dimensional linguistic processing in humans, computational benchmarks for cross-dimensional semantic inference, and longitudinal studies tracking the co-evolution of Cyberlanguage and the sociotechnical systems it mediates.

\section{Discussion}

\subsection{Theoretical Contributions}

Cyberlanguage makes contributions across several disciplinary domains. For linguistics, it extends inquiry from human natural language to heterogeneous-agent communication, advancing the field toward a general communication theory~\cite{shannon1948,clark1996,chomsky2002,manning1999,manovich2001}. For semiotics, the Cybersign extends Peirce's triadic structure to a quadruple framework for digital--physical hybrid environments~\cite{peirce1931,saussure1916,harnad1990,lakoff1980,aarseth1997}. For HCI, it reconstitutes the interaction problem as a multi-agent symbolic coordination problem embedded in four-dimensional space~\cite{suchman1987,dourish2001,licklider1960,norman2013,winograd1986,engelbart1962,galloway2012}. For philosophy of mind, it raises substantive questions about whether four-dimensional semantic content constitutes a natural kind~\cite{clark1997,dennett1991,zadeh1965,minsky1986,fodor1975}.

\subsection{Limitations and Critical Considerations}

The framework faces significant open challenges. The formalisation of the context-integration function~$\otimes$ remains underspecified; candidate designs range from learned weighted averaging to probabilistic graphical models, each with distinct failure modes. The cognitive feasibility of four-dimensional encoding for human users has not been established---it is possible that T-component handling will exceed working-memory constraints, requiring scaffolding. The governance of Cybersign mapping functions raises genuine ethical concerns: whoever designs the cross-dimensional mappings exercises normative power that must be made transparent and subject to democratic accountability~\cite{russell2020,bai2022,zhou2023,bender2021,gabriel2020,bender2020,floridi2019}.

A bootstrapping challenge also exists: Cyberlanguage cannot be specified without CyberCorpus, yet CyberCorpus cannot be collected without a specified Cyberlanguage. We resolve this circularity through the Phase~1 domain-specific strategy, where the annotation scheme is iteratively refined within bounded application domains before generalization.

\subsection{Practical Prospects}

In a 5--10 year horizon, the most tractable application domains are: smart-city crisis management (coordinating human responders, autonomous vehicles, AI decision-support, and digital twins); medical telemonitoring (linking patients, clinicians, diagnostic AIs, and wearable sensor streams); and human--robot collaborative manufacturing (integrating operator intent, workplace social conventions, robotic control systems, and real-time sensor data). Each domain provides a tractable testbed for the empirical predictions in Table~\ref{tab:predictions}~\cite{shi2025,rasheed2020,chen2024metaverse}.

\section{Conclusion}

This paper introduces Cyberlanguage as the native communication system of the CPST fusion space---an era in which physical environments, social organizations, cognitive processes, and digital infrastructures are deeply intertwined. We have argued that existing communication systems cannot meet the demands of this ontological condition, and proposed a theoretically grounded framework comprising the four-dimensional Cybersign semiotic model, Four-Dimensional Synchronous Grammar, a context-driven pragmatic mechanism, and a five-layer architectural stack. A staged implementation roadmap and four falsifiable empirical predictions constitute the immediate research agenda.

Cyberlanguage is proposed not as a replacement for natural language or programming languages, but as a meta-communicative infrastructure capable of coordinating heterogeneous agents---humans, AIs, robots, and digital entities---within an increasingly fused cyber--physical--social--cognitive reality. Its ultimate validation lies in empirical testability and real-world deployment.



\begin{thebibliography}{97}

\bibitem{wiener1961}
N.~Wiener. \textit{Cybernetics or Control and Communication in the Animal and the Machine}. MIT Press, 1961.

\bibitem{gerovitch2002}
S.~Gerovitch. \textit{From Newspeak to Cyberspeak: A History of Soviet Cybernetics}. MIT Press, 2002.

\bibitem{ning2026cyberism}
H.~Ning, J.~Ding, and K.~Michael. Cyberism: the fourth paradigm for the digital age. \textit{Computer}, 2026. DOI: 10.1109/MC.2026.3655852.

\bibitem{floridi2014}
L.~Floridi. \textit{The Fourth Revolution: How the Infosphere is Reshaping Human Reality}. Oxford University Press, 2014.

\bibitem{shannon1948}
C.~E.~Shannon and W.~Weaver. A mathematical theory of communication. \textit{The Bell System Technical Journal}, 27(3):379--423, 1948.

\bibitem{ning2025cybersophy}
H.~Ning. The fourth guiding principle for humans in the digital age: an initial exploration and framework construction of Cybersophy. \textit{Chinese Journal of Engineering}, 47(12):2470--2478, 2025.

\bibitem{ning2023cyberology}
H.~Ning et al. Cyberology: Cyber--Physical--Social-Thinking spaces-based discipline and interdiscipline hierarchy for metaverse (general cyberspace). \textit{IEEE Internet of Things Journal}, 10:4420--4430, 2023.

\bibitem{ning2015cpst}
H.~Ning and H.~Liu. Cyber-physical-social-thinking space based science and technology framework for the Internet of Things. \textit{Science China Information Sciences}, 58(3):1--19, 2015.

\bibitem{shi2025}
F.~Shi et al. National sovereignty and social governance based on cyber--physical--social--thinking space. \textit{Chinese Journal of Engineering}, 47(11):2257--2268, 2025.

\bibitem{stiegler1998}
B.~Stiegler. \textit{Technics and Time, 1: The Fault of Epimetheus}. Stanford University Press, 1998.

\bibitem{austin1962}
J.~L.~Austin. \textit{How to Do Things with Words}. Oxford University Press, 1962.

\bibitem{searle1969}
J.~R.~Searle. \textit{Speech Acts: An Essay in the Philosophy of Language}. Cambridge University Press, 1969.

\bibitem{gibson1979}
J.~J.~Gibson. \textit{The Ecological Approach to Visual Perception}. Houghton Mifflin, 1979.

\bibitem{gibson2015}
J.~J.~Gibson. \textit{The Ecological Approach to Visual Perception: Classic Edition}. Taylor \& Francis, 2015.

\bibitem{latour2005}
B.~Latour. \textit{Reassembling the Social: An Introduction to Actor-Network Theory}. Oxford University Press, 2005.

\bibitem{castells1996}
M.~Castells. \textit{The Rise of the Network Society}. Blackwell, 1996.

\bibitem{thrift2005}
N.~Thrift. \textit{Knowing Capitalism}. Sage, 2005.

\bibitem{hutchins1995}
E.~Hutchins. \textit{Cognition in the Wild}. MIT Press, 1995.

\bibitem{bateson2000}
G.~Bateson. \textit{Steps to an Ecology of Mind}. University of Chicago Press, 2000.

\bibitem{kitchin2011}
R.~Kitchin and M.~Dodge. \textit{Code/Space: Software and Everyday Life}. MIT Press, 2011.

\bibitem{hui2016}
Y.~Hui. \textit{On the Existence of Digital Objects}. University of Minnesota Press, 2016.

\bibitem{mcluhan1994}
M.~McLuhan. \textit{Understanding Media: The Extensions of Man}. MIT Press, 1994.

\bibitem{kittler1999}
F.~Kittler. \textit{Gramophone, Film, Typewriter}. Stanford University Press, 1999.

\bibitem{negroponte1995}
N.~Negroponte. \textit{Being Digital}. Knopf, 1995.

\bibitem{parikka2012}
J.~Parikka. \textit{What is Media Archaeology?} Polity Press, 2012.

\bibitem{bratton2015}
B.~H.~Bratton. \textit{The Stack: On Software and Sovereignty}. MIT Press, 2015.

\bibitem{benkler2006}
Y.~Benkler. \textit{The Wealth of Networks: How Social Production Transforms Markets and Freedom}. Yale University Press, 2006.

\bibitem{mitchell1995}
W.~J.~Mitchell. \textit{City of Bits: Space, Place, and the Infobahn}. MIT Press, 1995.

\bibitem{willett2023}
F.~R.~Willett et al. A high-performance speech neuroprosthesis. \textit{Nature}, 620:1031--1036, 2023.

\bibitem{metzger2023}
S.~L.~Metzger et al. A high-performance neuroprosthesis for speech decoding and avatar control. \textit{Nature}, 620:1037--1046, 2023.

\bibitem{rasheed2020}
A.~Rasheed et al. Digital twins: values, challenges and enablers from a modeling perspective. \textit{IEEE Access}, 8:21980--22012, 2020.

\bibitem{clark2003}
A.~Clark. \textit{Natural-Born Cyborgs: Minds, Technologies, and the Future of Human Intelligence}. Oxford University Press, 2003.

\bibitem{hayles1999}
N.~K.~Hayles. \textit{How We Became Posthuman: Virtual Bodies in Cybernetics, Literature, and Informatics}. University of Chicago Press, 1999.

\bibitem{haraway1991}
D.~Haraway. A cyborg manifesto. In \textit{Simians, Cyborgs and Women: The Reinvention of Nature}, pages 149--181. Routledge, 1991.

\bibitem{hansen2006}
M.~B.~N.~Hansen. \textit{Bodies in Code: Interfaces with Digital Media}. Routledge, 2006.

\bibitem{peirce1931}
C.~S.~Peirce. \textit{Collected Papers of Charles Sanders Peirce}, Vols.~1--8. Harvard University Press, 1931--1958.

\bibitem{saussure1916}
F.~de~Saussure. \textit{Course in General Linguistics}. Open Court, 1916/1983.

\bibitem{clark1996}
H.~H.~Clark. \textit{Using Language}. Cambridge University Press, 1996.

\bibitem{brooks1991}
R.~A.~Brooks. Intelligence without representation. \textit{Artificial Intelligence}, 47:139--159, 1991.

\bibitem{park2023}
J.~S.~Park et al. Generative agents: interactive simulacra of human behavior. In \textit{Proc.\ 36th Annual ACM Symposium on User Interface Software and Technology}, pages 1--22. ACM, 2023.

\bibitem{yao2023}
S.~Yao et al. ReAct: synergizing reasoning and acting in language models. In \textit{Proc.\ ICLR 2023}. OpenReview, 2023.

\bibitem{zhao2023}
W.~X.~Zhao et al. A survey of large language models. Preprint at arXiv, 2023. \url{https://doi.org/10.48550/arXiv.2303.18223}.

\bibitem{xi2023}
Z.~Xi et al. The rise and potential of large language model based agents: a survey. Preprint at arXiv, 2023. \url{https://doi.org/10.48550/arXiv.2309.07864}.

\bibitem{touvron2023}
H.~Touvron et al. LLaMA: open and efficient foundation language models. Preprint at arXiv, 2023. \url{https://doi.org/10.48550/arXiv.2302.13971}.

\bibitem{pan2024}
X.~Pan and O.~Schwartz. Multimodal AI needs active human interaction. \textit{Nature Human Behaviour}, 8(10):1825--1826, 2024.

\bibitem{li2023camel}
G.~Li et al. CAMEL: communicative agents for `mind' exploration of large language model society. In \textit{Advances in Neural Information Processing Systems}, 36, 2023.

\bibitem{wang2024}
L.~Wang et al. A survey on large language model based autonomous agents. \textit{Frontiers of Computer Science}, 18:186345, 2024.

\bibitem{xie2021}
H.~Xie et al. Deep learning enabled semantic communication systems. \textit{IEEE Journal on Selected Areas in Communications}, 39:2253--2270, 2021.

\bibitem{qin2022}
Z.~Qin et al. Semantic communications: principles and challenges. Preprint at arXiv, 2022. \url{https://doi.org/10.48550/arXiv.2212.00032}.

\bibitem{gunduz2023}
D.~G\"{u}nd\"{u}z et al. Beyond transmitting bits: context, semantics, and task-oriented communications. \textit{IEEE Journal on Selected Areas in Communications}, 41:5--41, 2023.

\bibitem{harnad1990}
S.~Harnad. The symbol grounding problem. \textit{Physica D}, 42:335--346, 1990.

\bibitem{lakoff1980}
G.~Lakoff and M.~Johnson. \textit{Metaphors We Live By}. University of Chicago Press, 1980.

\bibitem{wittgenstein1953}
L.~Wittgenstein. \textit{Philosophical Investigations}. Blackwell, 1953.

\bibitem{turing1950}
A.~M.~Turing. Computing machinery and intelligence. \textit{Mind}, 59:433--460, 1950.

\bibitem{clark1997}
A.~Clark. \textit{Being There: Putting Brain, Body, and World Together Again}. MIT Press, 1997.

\bibitem{dennett1991}
D.~C.~Dennett. \textit{Consciousness Explained}. Little, Brown, 1991.

\bibitem{chomsky2002}
N.~Chomsky. \textit{Syntactic Structures}. Mouton de Gruyter, 2002.

\bibitem{manning1999}
C.~D.~Manning and H.~Sch\"{u}tze. \textit{Foundations of Statistical Natural Language Processing}. MIT Press, 1999.

\bibitem{newell1976}
A.~Newell and H.~A.~Simon. Computer science as empirical inquiry: symbols and search. \textit{Communications of the ACM}, 19:113--126, 1976.

\bibitem{grice1978}
H.~P.~Grice. Further notes on logic and conversation. In \textit{Syntax and Semantics}, Vol.~9, pages 113--127, 1978.

\bibitem{pan2024kg}
J.~Z.~Pan et al. Unifying large language models and knowledge graphs: a roadmap. \textit{IEEE Transactions on Knowledge and Data Engineering}, 36:3580--3599, 2024.

\bibitem{hogan2021}
A.~Hogan et al. Knowledge graphs. \textit{ACM Computing Surveys}, 54:71, 2021.

\bibitem{bernerslee2001}
T.~Berners-Lee, J.~Hendler, and O.~Lassila. The semantic web. \textit{Scientific American}, 284:34--43, 2001.

\bibitem{kress2001}
G.~Kress and T.~van~Leeuwen. \textit{Multimodal Discourse: The Modes and Media of Contemporary Communication}. Arnold, 2001.

\bibitem{tomasello2008}
M.~Tomasello. \textit{Origins of Human Communication}. MIT Press, 2008.

\bibitem{shumailov2024}
I.~Shumailov et al. AI models collapse when trained on recursively generated data. \textit{Nature}, 631:755--759, 2024.

\bibitem{wei2022emergent}
J.~Wei et al. Emergent abilities of large language models. \textit{Transactions on Machine Learning Research}, 2022.

\bibitem{bommasani2022}
R.~Bommasani et al. On the opportunities and risks of foundation models. Preprint at arXiv, 2022. \url{https://doi.org/10.48550/arXiv.2108.07258}.

\bibitem{achiam2023}
J.~Achiam et al. GPT-4 technical report. Preprint at arXiv, 2023. \url{https://doi.org/10.48550/arXiv.2303.08774}.

\bibitem{gu2024}
A.~Gu and T.~Dao. Mamba: linear-time sequence modeling with selective state spaces. In \textit{Proc.\ ICLR 2024}. OpenReview, 2024.

\bibitem{wei2022cot}
J.~Wei et al. Chain-of-thought prompting elicits reasoning in large language models. In \textit{Advances in Neural Information Processing Systems}, 35, 2022.

\bibitem{vaswani2017}
A.~Vaswani et al. Attention is all you need. In \textit{Advances in Neural Information Processing Systems}, 30, 2017.

\bibitem{brown2020}
T.~Brown et al. Language models are few-shot learners. In \textit{Advances in Neural Information Processing Systems}, 33:1877--1901, 2020.

\bibitem{lecun2015}
Y.~LeCun, Y.~Bengio, and G.~Hinton. Deep learning. \textit{Nature}, 521:436--444, 2015.

\bibitem{silver2016}
D.~Silver et al. Mastering the game of Go with deep neural networks and tree search. \textit{Nature}, 529:484--489, 2016.

\bibitem{russell2020}
S.~J.~Russell and P.~Norvig. \textit{Artificial Intelligence: A Modern Approach}, 4th edn. Pearson, 2020.

\bibitem{levy1997}
P.~L\'{e}vy. \textit{Collective Intelligence: Mankind's Emerging World in Cyberspace}. Plenum Press, 1997.

\bibitem{lessig1999}
L.~Lessig. \textit{Code and Other Laws of Cyberspace}. Basic Books, 1999.

\bibitem{manovich2001}
L.~Manovich. \textit{The Language of New Media}. MIT Press, 2001.

\bibitem{aarseth1997}
E.~J.~Aarseth. \textit{Cybertext: Perspectives on Ergodic Literature}. Johns Hopkins University Press, 1997.

\bibitem{suchman1987}
L.~Suchman. \textit{Plans and Situated Actions: The Problem of Human-Machine Communication}. Cambridge University Press, 1987.

\bibitem{dourish2001}
P.~Dourish. \textit{Where the Action Is: The Foundations of Embodied Interaction}. MIT Press, 2001.

\bibitem{licklider1960}
J.~C.~R.~Licklider. Man-computer symbiosis. \textit{IRE Transactions on Human Factors in Electronics}, HFE-1:4--11, 1960.

\bibitem{norman2013}
D.~A.~Norman. \textit{The Design of Everyday Things}. MIT Press, 2013.

\bibitem{winograd1986}
T.~Winograd and F.~Flores. \textit{Understanding Computers and Cognition: A New Foundation for Design}. Ablex, 1986.

\bibitem{engelbart1962}
D.~C.~Engelbart. Augmenting human intellect: a conceptual framework. Tech.\ Rep.\ AFOSR-3233, Stanford Research Institute, 1962.

\bibitem{galloway2012}
A.~R.~Galloway. \textit{The Interface Effect}. Polity Press, 2012.

\bibitem{zadeh1965}
L.~A.~Zadeh. Fuzzy sets. \textit{Information and Control}, 8:338--353, 1965.

\bibitem{minsky1986}
M.~Minsky. \textit{The Society of Mind}. Simon \& Schuster, 1986.

\bibitem{fodor1975}
J.~A.~Fodor. \textit{The Language of Thought}. Crowell, 1975.

\bibitem{bai2022}
Y.~Bai et al. Constitutional AI: harmlessness from AI feedback. Preprint at arXiv, 2022. \url{https://doi.org/10.48550/arXiv.2212.08073}.

\bibitem{zhou2023}
C.~Zhou et al. LIMA: less is more for alignment. In \textit{Advances in Neural Information Processing Systems}, 36, 2023.

\bibitem{bender2021}
E.~M.~Bender et al. On the dangers of stochastic parrots: can language models be too big? In \textit{Proc.\ FAccT 2021}, pages 610--623. ACM, 2021.

\bibitem{gabriel2020}
I.~Gabriel. Artificial intelligence, values, and alignment. \textit{Minds and Machines}, 30:411--437, 2020.

\bibitem{bender2020}
E.~M.~Bender and A.~Koller. Climbing towards NLU: on meaning, form, and understanding in the age of data. In \textit{Proc.\ ACL 2020}, pages 5185--5198, 2020.

\bibitem{floridi2019}
L.~Floridi et al. An ethical framework for a good AI society: opportunities, risks, principles, and recommendations. \textit{Minds and Machines}, 29:689--707, 2019.

\bibitem{chen2024metaverse}
Z.~Chen et al. Metaverse for smart cities: a survey. \textit{Internet of Things and Cyber-Physical Systems}, 4:203--216, 2024.

\end{thebibliography}
\end{document}